\documentclass[twocolumn,aps,pra,amsmath,floatfix]{revtex4}
\usepackage{graphicx}
\begin{document}

\title{Surface versus bulk Mott transition in Ca$_x$La$_{1-x}$VO$_3$}

\author{A. Liebsch}
\affiliation{Institut f\"ur Festk\"orperforschung, Forschungszentrum
             J\"ulich, 52425 J\"ulich, Germany}
\date{\today }

\begin{abstract}
The Mott insulator LaVO$_3$ is known to become metallic if La is 
replaced by small concentrations of Ca. Since surface quasi-particle 
spectra of various perovskites are more strongly correlated than 
their bulk spectra, the coexistence of a metallic bulk and an 
insulating surface layer might be feasible. To investigate this 
possibility the dynamical mean field theory is used to evaluate the 
quasi-particle spectra in the bulk and at the surface of 
Ca$_x$La$_{1-x}$VO$_3$ for various Ca concentrations.
\end{abstract}
\maketitle

Significant surface contributions to quasi-particle spectra of 
strongly correlated transition metal oxides have recently been 
identified in photoemission measurements on several perovskite materials:
SrRuO$_3$\ \cite{fujioka},
Sr$_2$RuO$_4$\ \cite{SrRuO1,SrRuO2},
Ca$_x$Sr$_{1-x}$VO$_3$\ \cite{aiura,maiti,sekiyama}, and
Ca$_x$La$_{1-x}$VO$_3$\ \cite{maiti00}.
The latter series is particularly interesting since LaVO$_3$ is
insulating and becomes metallic upon doping with Ca. By varying the 
photon energy Maiti {\it et al.}\ \cite{maiti00} were able to show 
that surface correlation features are more pronounced than in the bulk, 
in accord with the observations on the other perovskites cited above. 
Thus, the coherent peak at the Fermi level is less intense and the 
satellite peak below the V $t_{2g}$ band is stronger than observed 
in the corresponding bulk photoemission spectra. Even more remarkably, 
Maiti {\it et al.} identified a Ca concentration range where the 
bulk has become metallic while the surface still appears insulating, 
before, at even higher Ca doping, it also becomes metallic. 
Although sample preparation and experimental resolution make the 
data analysis difficult, these results address a very interesting 
issue, namely the possible coexistence of different bulk and surface 
phases in semi-infinite systems.
In the case of magnetism, for example, it is well known that at finite
temperatures the magnetic moment can be reduced in the surface layers 
of a ferromagnet, but that both have the same Curie temperature
unless the surface is artificially decoupled from the bulk  
\ \cite{magnetism}. 

The identification surface-induced enhanced correlations is clearly
important in order not to interprete them erroneously as bulk
correlations. For instance, earlier photoemission data on 
Ca$_x$Sr$_{1-x}$VO$_3$\ \cite{aiura} appeared to be in conflict with 
various low-frequency measurements\ \cite{onada}. While the latter 
data showed that this perovskite series is metallic for all Ca 
concentrations, the photoemission spectra seemed to suggest that 
SrVO$_3$ is strongly correlated and that CaVO$_3$ is on the verge 
to a metal-insulator transition. Recent photoemission measurements 
\ \cite{maiti,sekiyama} covering a wide range of photon energies
demonstrated instead that the bulk spectra of both materials are 
consistent with metallic behavior, but that correlation features
are much more pronounced at the surface. While SrVO$_3$ is a cubic
perovskite, the oxygen octahedra in CaVO$_3$ are orthorhombically
distorted. This leads to a slight reduction in $t_{2g}$ band width
and a broadening of the van Hove singularity. The density of states 
at $E_F$, however, is only weakly affected. That the bulk properties 
of SrVO$_3$ and CaVO$_3$ should indeed be very similar was recently 
also found by Nekrasov {\it et al.} \cite{nekrasov} in bulk electronic 
structure calculations combining the the local density approximation 
(LDA) with the many-body dynamical mean field theory (DMFT)\ 
\cite{georges,vollhardt,pruschke}. 

As shown by Liebsch\ \cite{lieprl} using surface DMFT calculations,  
the enhancement of surface correlations can at least in part be 
understood in terms of the reduced atomic coordination at the surface 
and the concomittant narrowing of the local density of states. 
Thus, the ratio of local Coulomb energy and band width $U/W$ is
effectively enhanced at the surface. Accordingly, the coherent 
peak in surface quasi-particle spectra is less intense and the 
lower Hubbard band is enhanced compared to the bulk. Other 
mechanisms, such as additional band narrowing due to reconstruction
and distortion of oxygen octahedra, or less efficient screening of 
electron-electron interactions at surfaces, might further increase
the importance of correlations in surface photoemission spectra.
The recent work on Ca$_x$Sr$_{1-x}$VO$_3$ therefore established
that photoemission is consistent with low-frequency bulk probes 
as long as surface effects are properly taken into account.

The same conclusion applies to Sr$_2$RuO$_4$ for which previous 
photoemission spectra seemed to contradict bulk de Haas-van Alphen 
measurements\ \cite{SrRuO1,SrRuO3}. Recent experimental and 
theoretical work\ \cite{SrRuO2,SrRuO4} proved, however, that this 
discrepancy can be resolved by taking into account the lattice 
reconstruction at the surface of Sr$_2$RuO$_4$ which leads to 
significant changes in the photoemission spectra.
  
In the present work we investigate surface correlations in the
perovskite series Ca$_x$Sr$_{1-x}$VO$_3$ at Ca doping levels where 
metallic bulk and insulating surface behavior might coexist.
Since doping with Ca leads only to minor structural changes and 
nearly constant Hubbard $U$\ \cite{bocquet}, the modifications 
of the electronic properties are almost entirely caused by the 
degree of band filling. According to theoretical predictions by 
Potthoff and Nolting\ \cite{potthoff}, a phase with metallic bulk
and insulating surface properties cannot occur unless
electronic coupling between surface and bulk is assumed to be weak.
These results were derived for the semi-infinite half-filled simple 
cubic $s$ band, and by adopting a linearlized version of the DMFT in 
which low- and high-frequency excitations are essentially decoupled.     
It is therefore not clear to what extent they are applicable to 
realistic 
multi-band systems at arbitrary band filling. Here we calculate
surface quasi-particle spectra by using a realistic layer-dependent 
local density of states appropriate for Ca$_x$Sr$_{1-x}$VO$_3$, 
combined with multi-band DMFT Quantum Monte Carlo (QMC) calculations.

\begin{figure}[t!]%1
  \begin{center}
  \includegraphics[width=4.5cm,height=7.5cm,angle=-90]{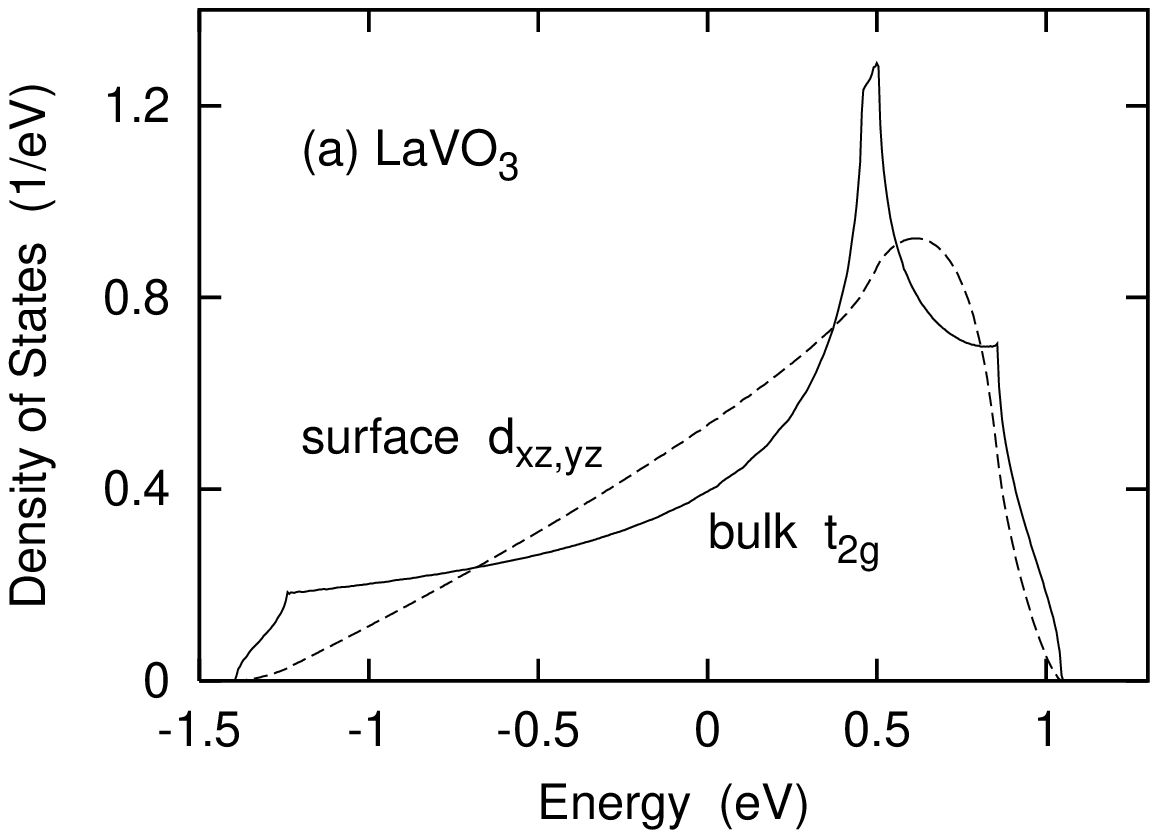}
  \includegraphics[width=4.5cm,height=7.5cm,angle=-90]{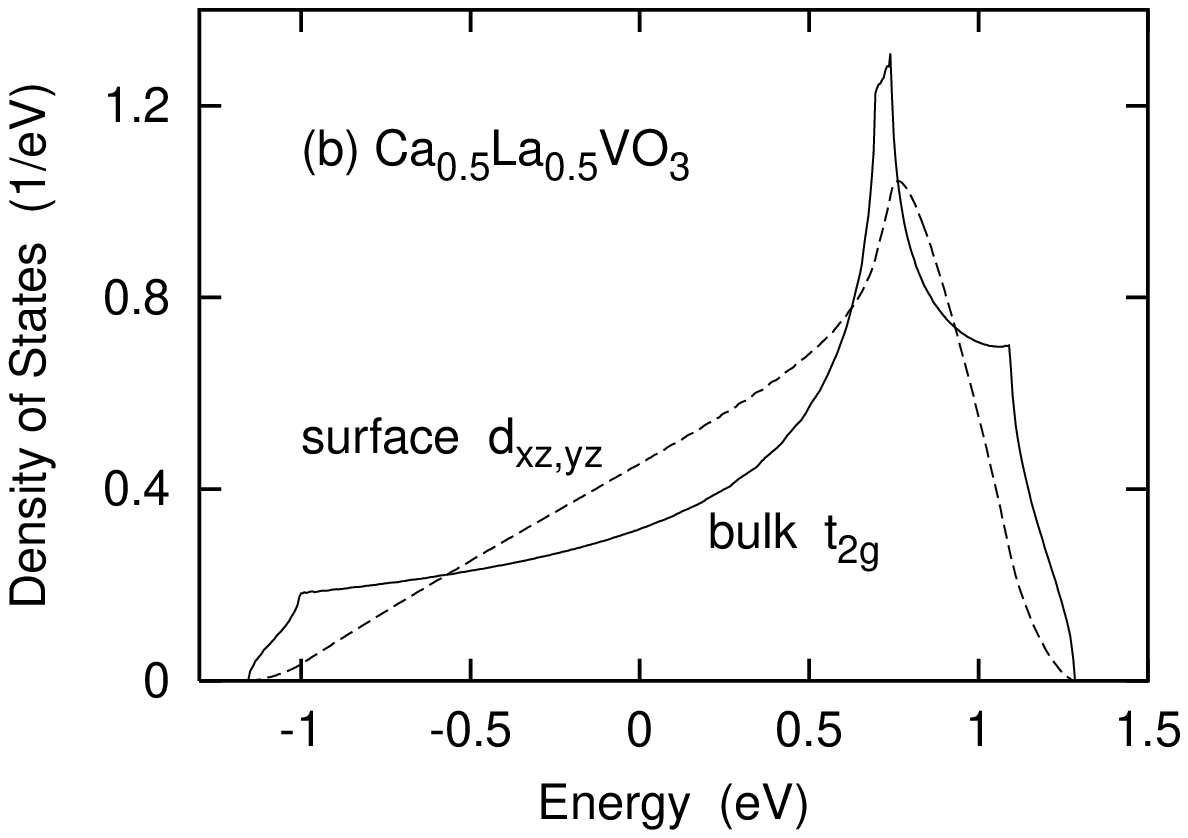}
  \end{center}
\caption{
Solid curves: isotropic bulk density of states of cubic (a) LaVO$_3$ 
($d^2$) and (b) Ca$_{0.5}$La$_{0.5}$VO$_3$ ($d^{1.5}$). Dashed curves: 
local density of $d_{xz,yz}$ states in the first layer ($E_F=0$).
}\end{figure}

Fig.~1 shows the bulk and surface density of states for LaVO$_3$
and Ca$_{0.5}$La$_{0.5}$VO$_3$. The main difference is the band
filling. While the $d^2$ occupancy of the V $t_{2g}$ bands makes 
LaVO$_3$ insulating\ \cite{solovyev}, Ca$_{0.5}$La$_{0.5}$VO$_3$ 
with its  $d^{1.5}$ occupancy is metallic. In the bulk, the 
insulator-metal transition occurs at about $x\approx0.2$\ 
\cite{maiti}. Since our main interest here is to study the effect 
of band filling on this transition in the bulk and 
at the surface we adopt for simplicity the ideal cubic symmetry 
and neglect the distortion of oxygen octahedra. According to the 
LDA-DMFT calculations for CaVO$_3$\ \cite{nekrasov} these structural 
changes have little influence on the $t_{2g}$ quasi-particle spectra.
The bulk density of states is therefore similar to that of SrVO$_3$. 
It exhibits the characteristic asymmetric shape resulting from the 
pronounced planar nature of the $t_{2g}$ states which is found for 
various peroskite materials\ \cite{takegahara}. 

In a cubic structure, the $t_{2g}$ bands are degenerate. The 
surface lifts this degeneracy since only the intra-planar 
$d_{xy}$ states exhibit strong dispersion within the surface 
layer while $d_{xz,yz}$ bands are modified due to the reduced
atomic coordination (the $z$-axis specifies the surface normal).
The layer dependent local density of states is calculated using
a tight-binding formalism for semi-infinite systems\ \cite{kalkstein}.
Details concerning the application of this method to $t_{2g}$ bands
are given Ref.\ \cite{lieprb}. 
As shown in Fig.~1, the spectral weight is shifted from the low-
and high-energy regions to intermediate energies close to $E_F$.
Thus, although the full band width coincides with the one in the
bulk, the local density of $d_{xz,yz}$ states in the surface layer is 
effectively narrowed. The surface distributions shown in Fig.~1\,(a) 
and (b) are qualitatively similar. The small difference arises from
slightly different surface potentials which are employed in the 
electronic structure calculation for semi-infinite
Ca$_x$La$_{1-x}$VO$_3$ in order to ensure charge neutrality.
In the deeper layers, the local density of $d_{xz,yz}$ states
approaches the bulk $t_{2g}$  density rather quickly, the main
modification consisting in an oscillatory deviation which 
decreases with increasing distance from the surface.  
In contrast to the $d_{xz,yz}$ densities, the local density of 
$d_{xy}$ states in the surface layer is almost identical to the 
isotropic bulk density. This is a consequence of the pronounced 
planar character of the $t_{2g}$ states. 
 
The narrowing of the local density states in the surface layer 
has an important influence on the quasi-particle distributions
which we evaluate by using the multiband QMC-DMFT. Essentially
we are dealing with coupled impurity problems where layer-dependent
self-energies must be calculated and iterated until self-consistency
is achieved. Since the full solution of this problem for multiband 
systems is computationally not yet feasible, we neglect the coupling 
between layer baths and treat bulk and surface layers independently. 
In previous single-band DMFT calculations for surfaces this coupling  
was found to be small\ \cite{potthoff}. Enforcing charge neutrality 
at the surface presumably also diminishes the importance of 
inter-layer coupling between the impurity baths. The key input in
the DMFT calculation is therefore the orbital- and layer-dependent 
local density of states. As a result of the one-electron band narrowing
we can expect a similar enhancement of surface correlation effects 
as discussed previously for SrVO$_3$ and SrRuO$_3$\ \cite{lieprb}.

\begin{figure}[t!]%2
\begin{center}     
    \includegraphics[height=7cm,width=4.0cm,angle=-90]{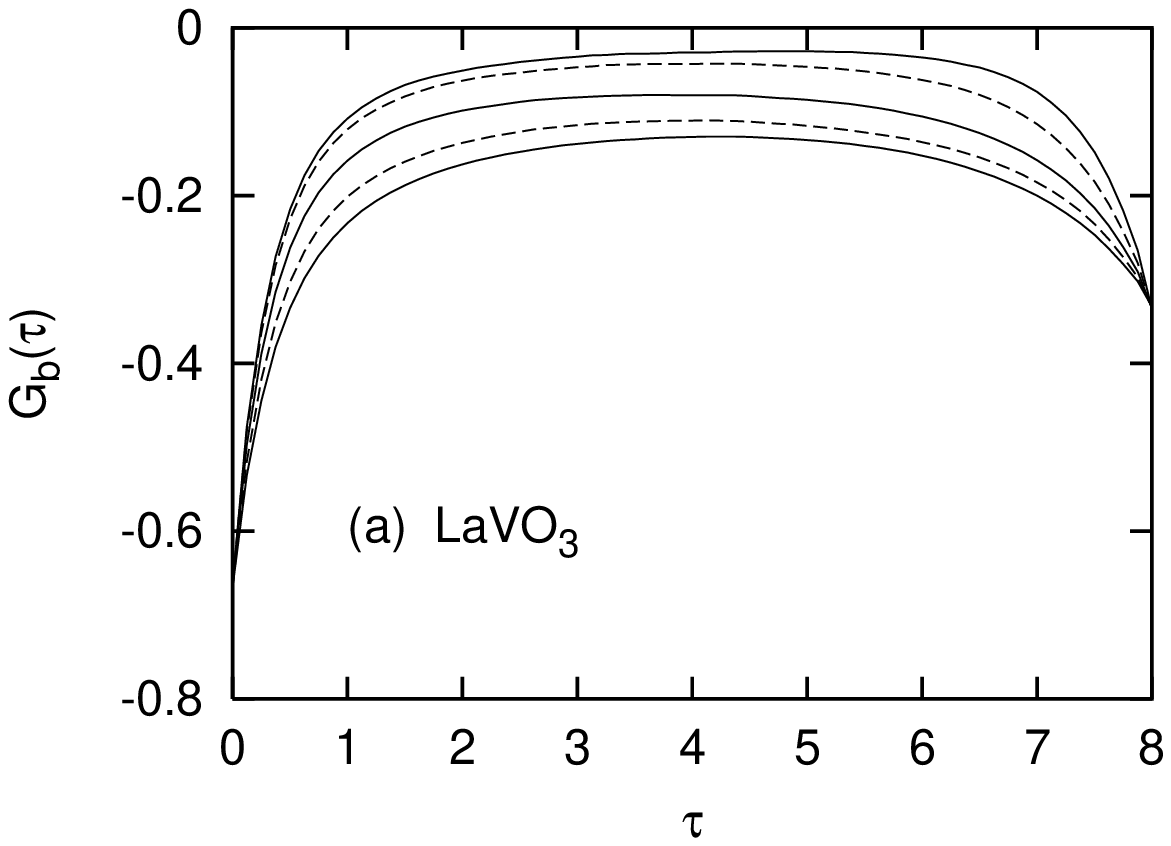}
    \includegraphics[height=7cm,width=4.0cm,angle=-90]{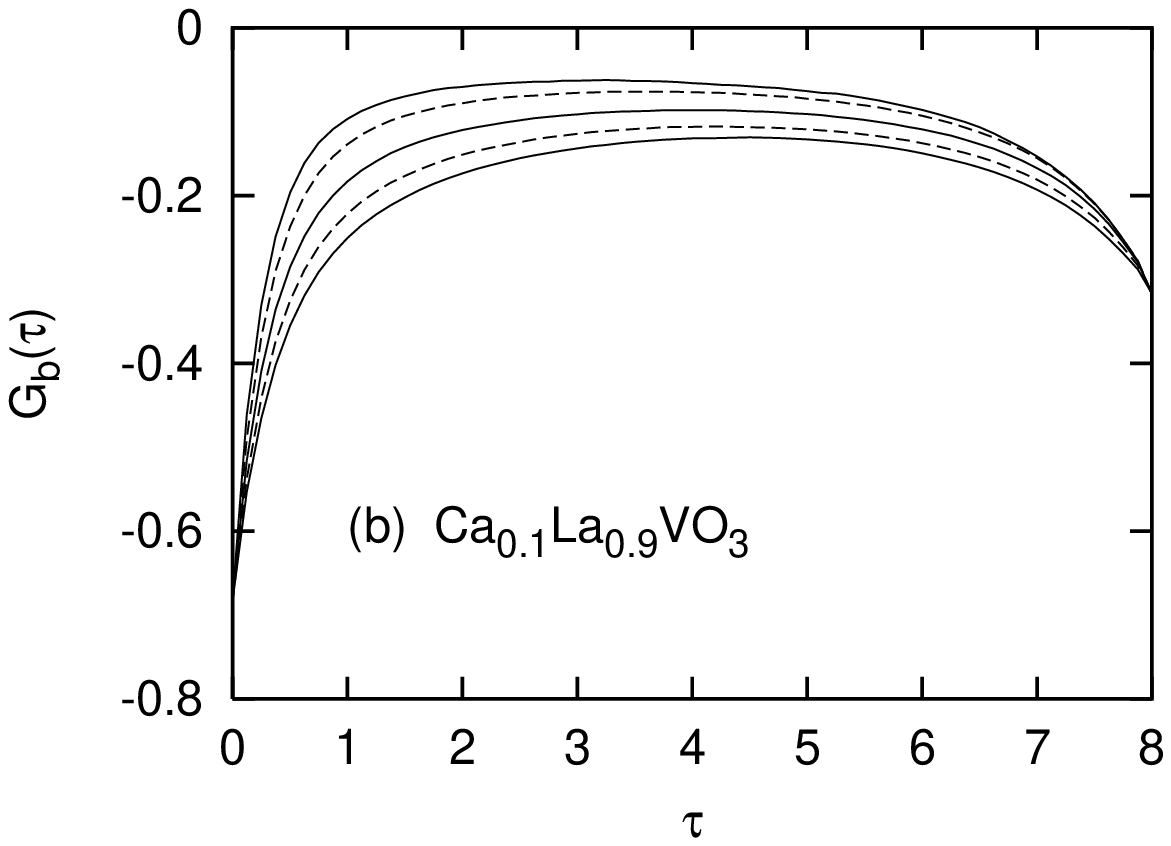}
    \includegraphics[height=7cm,width=4.0cm,angle=-90]{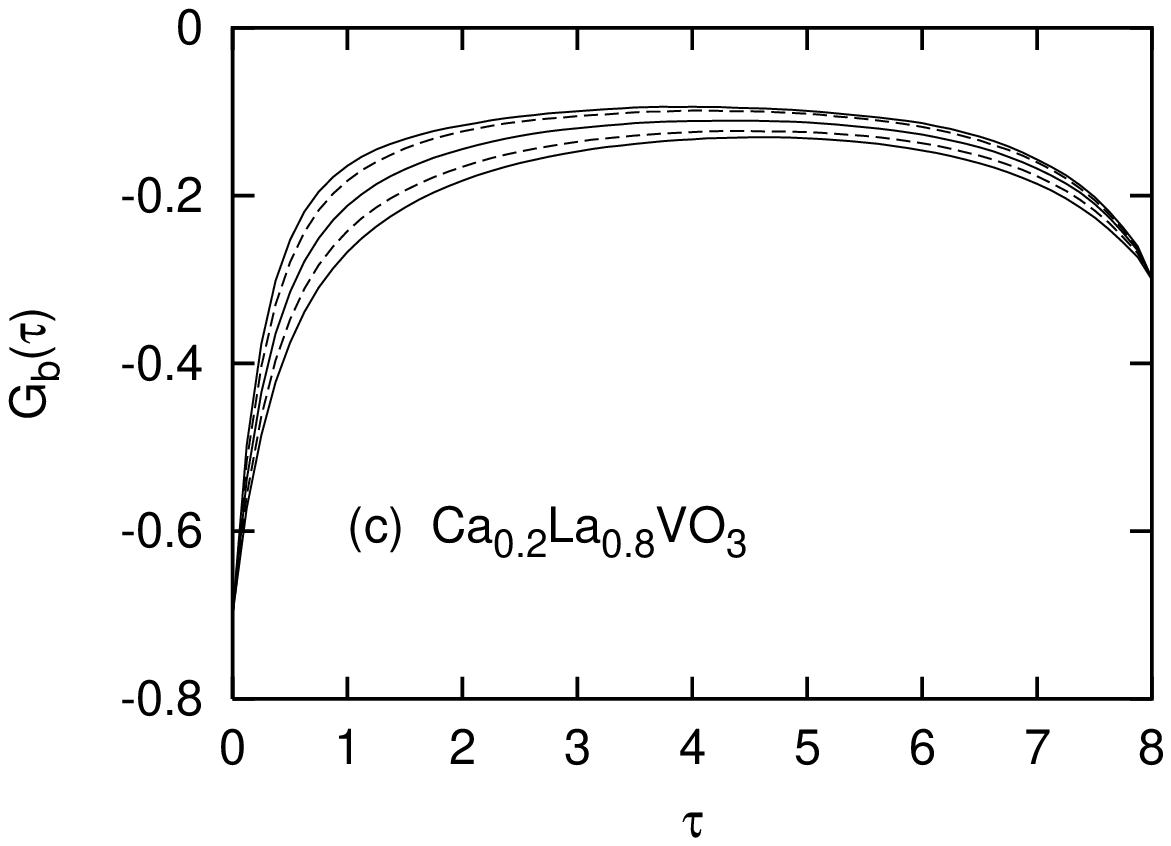}
    \includegraphics[height=7cm,width=4.0cm,angle=-90]{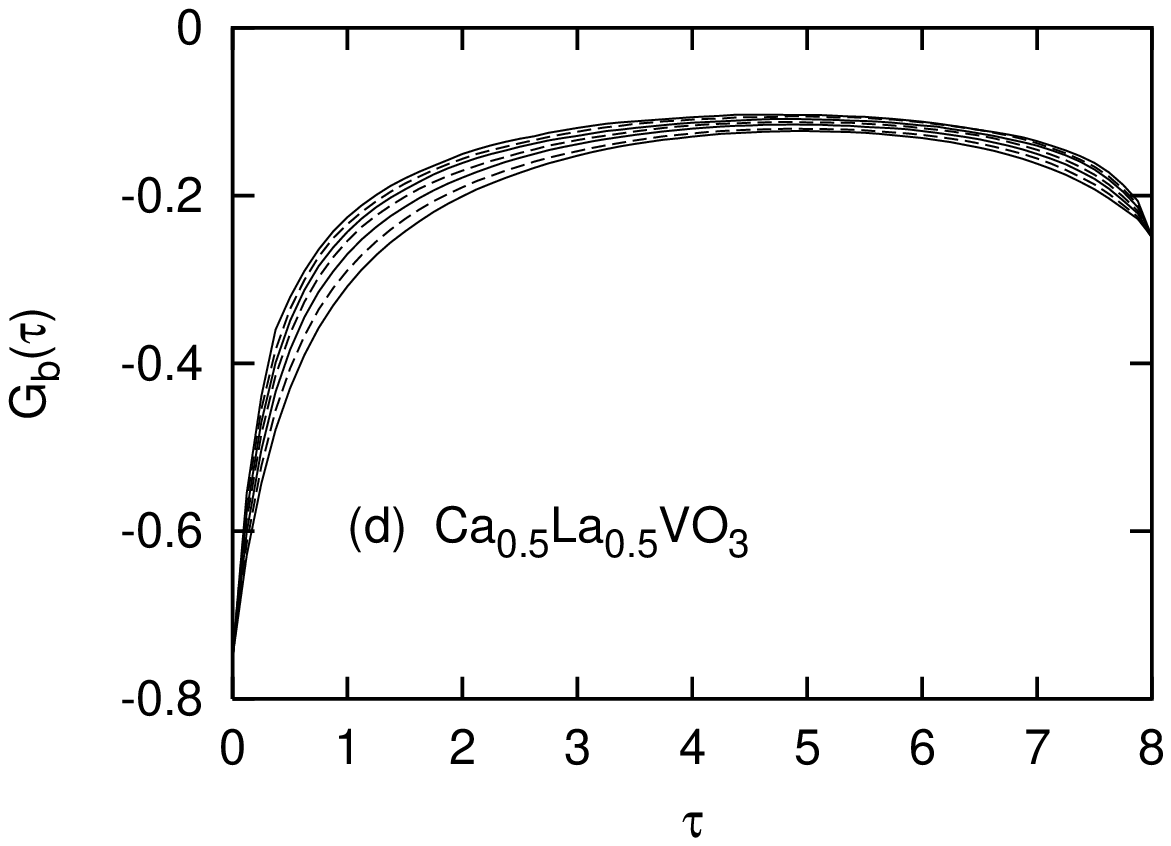}
\end{center}
\caption{
Quasi-particle Green's function as function of imaginary time 
for bulk $t_{2g}$ states of Ca$_x$La$_{1-x}$VO$_3$ derived from DMFT 
($\beta=8$). $G(\beta/2)$ is representative of the quasi-particle 
weight at $E_F$.
(a) $x=0$; (b) $x=0.1$; (c) $x=0.2$; (d) $x=0.5$.
Plotted are $G(\tau)$ for $U=3.0\ldots5.0$~eV [up to 6~eV in (d)]
in steps of 0.5~eV (from below); $J=0.7$~eV.
}\end{figure}
 
As mentioned above, LaVO$_3$ is insulating, but hole doping by 
replacement of La via Ca induces an insulator-metal transition at 
rather low Ca concentrations. This can be seen clearly in the 
different shapes of the imaginary-time Green's function for bulk  
Ca$_{x}$La$_{1-x}$VO$_3$, as shown in Fig.~2 for several values of 
$x$. Note that the value 
of $G(\tau)$ at $\tau\approx\beta/2$ is representative of the weight 
of the quasi-particle peak at $E_F$, while $G(\beta)$ gives the 
occupancy per spin-band. In the case of LaVO$_3$ ($x=0$) already for 
$U=4.5\ldots5.0$~eV there is little weight near $\tau\approx\beta/2$, 
indicating the reduction of intensity near $E_F$ as $U$ approaches 
the critical value for the Mott transition. Of course, since the
DMFT-QMC calculations are carried out at finite temperatures, 
$G(\beta/2)$ cannot approach zero, i.e., a finite spectral density
at $E_F$ remains even in the nominally insulating phase. According
to the results shown at finite Ca concentrations, however, 
$G(\beta/2)$ diminishes much less rapidly, indicating the 
incipient metallicity for $x>0$. Thus, in the range of Coulomb
energies where LaVO$_3$ is insulating, i.e., for $U\approx5.0$~eV,  
Ca$_{x}$La$_{1-x}$VO$_3$ for $x>0.2$ is metallic. We conclude from 
these calculations that within our model the borderline doping level
between insulating and metallic behavior is about $x\approx0.1$. 
 
\begin{figure}[t!]%3
\begin{center}     
    \includegraphics[height=7cm,width=4.0cm,angle=-90]{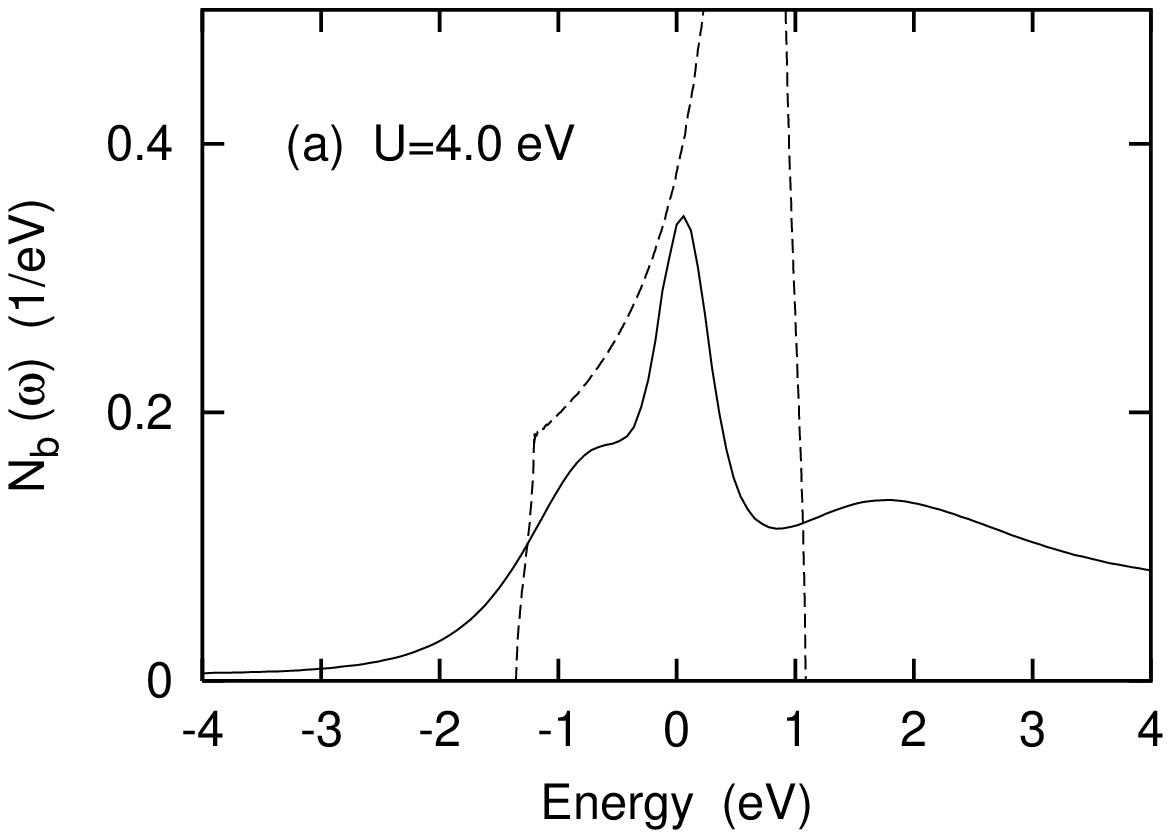}
    \includegraphics[height=7cm,width=4.0cm,angle=-90]{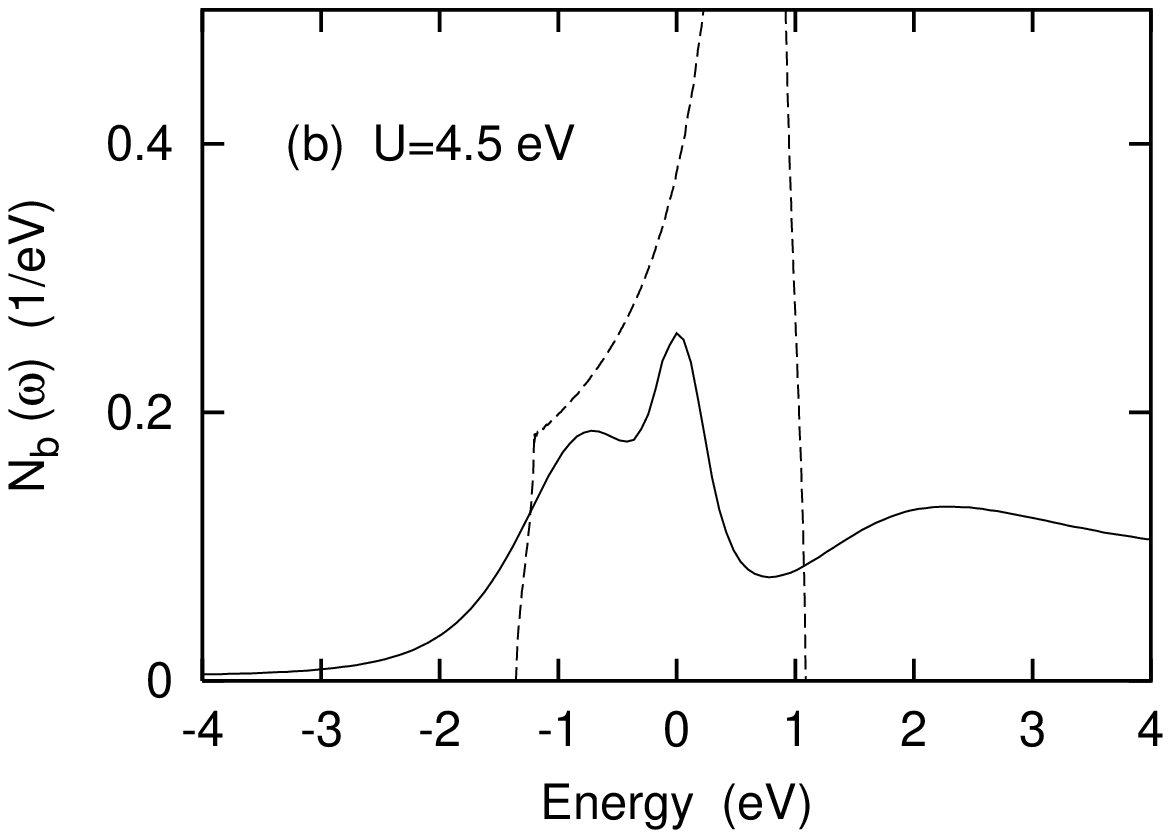}
    \includegraphics[height=7cm,width=4.0cm,angle=-90]{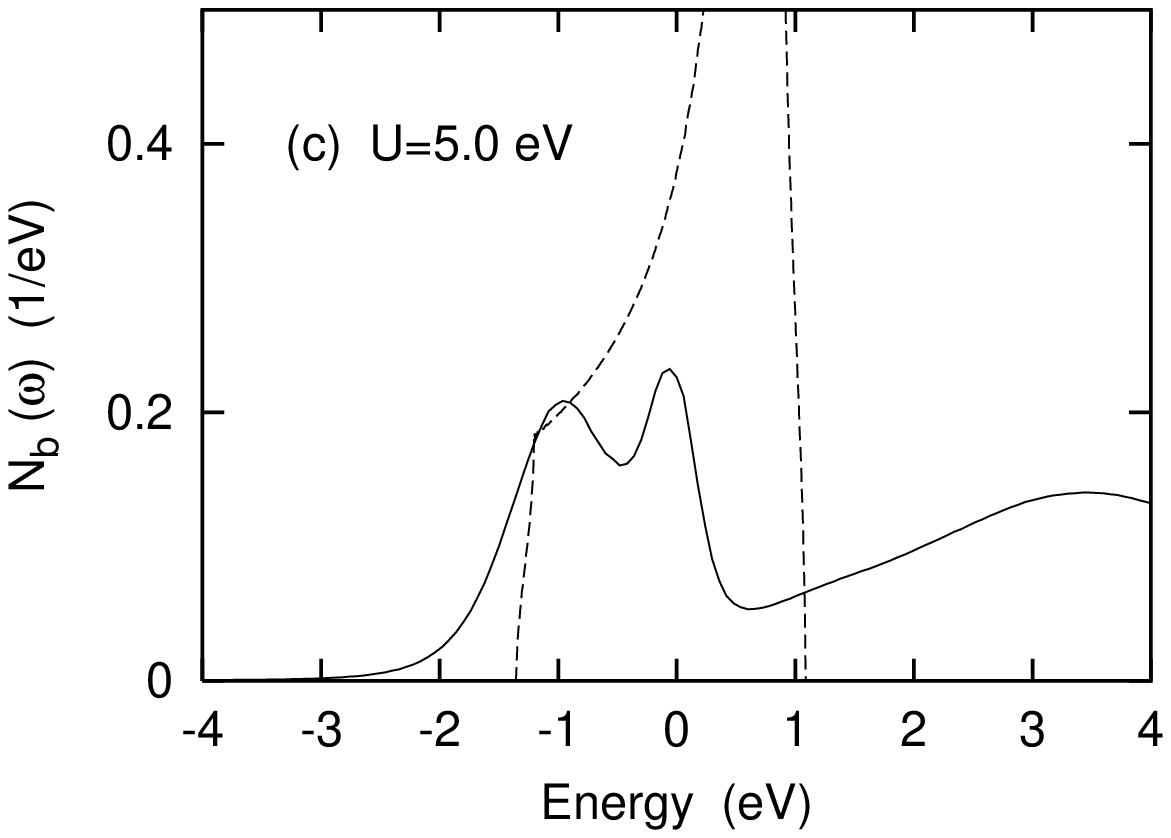}
\end{center}%vskip16cm
\caption{
Quasi-particle density of bulk $t_{2g}$ states of 
Ca$_{0.1}$La$_{0.9}$VO$_3$ ($d^{1.9}$) for (a) $U=4.0$~eV, (b) 4.5~eV
and (c) 5.0~eV derived from DMFT (solid curves); $J=0.7$~eV. 
Dashed curves: bare density of states. 
}\end{figure}

\begin{figure}[t!]%4
\begin{center}     
    \includegraphics[height=7cm,width=4.0cm,angle=-90]{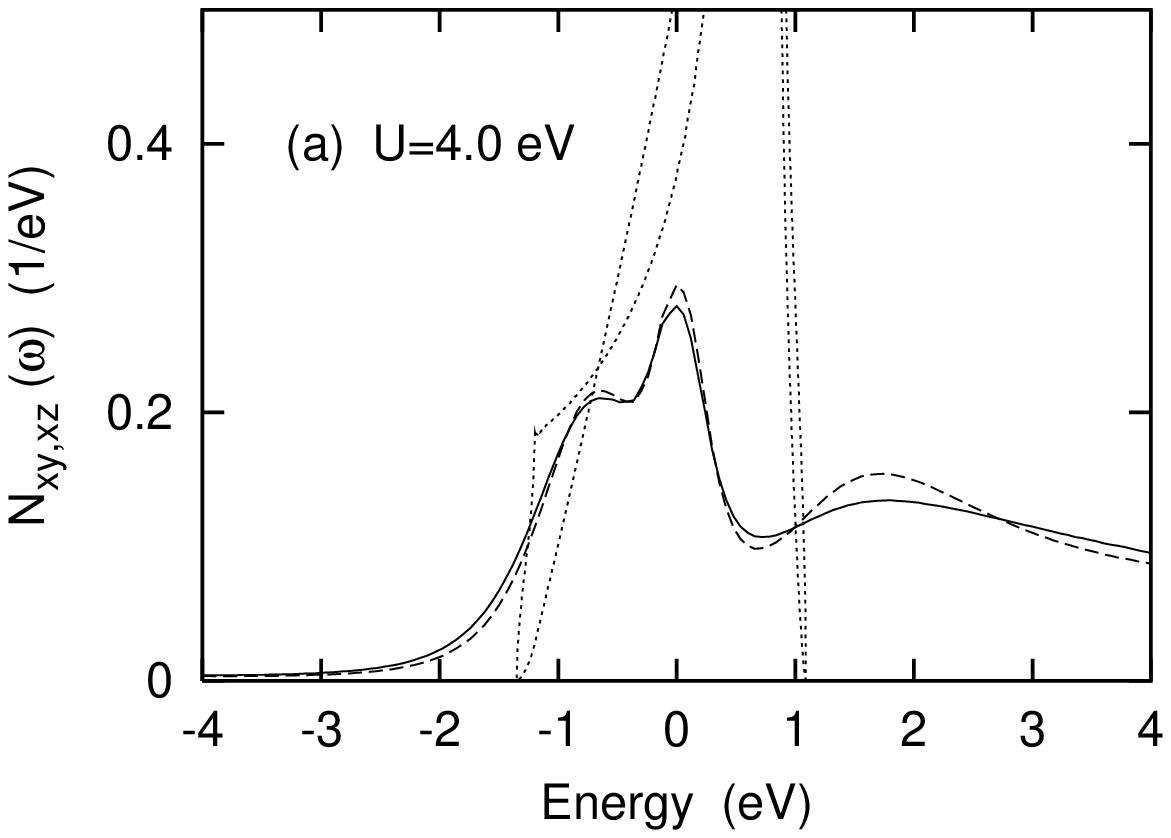}
    \includegraphics[height=7cm,width=4.0cm,angle=-90]{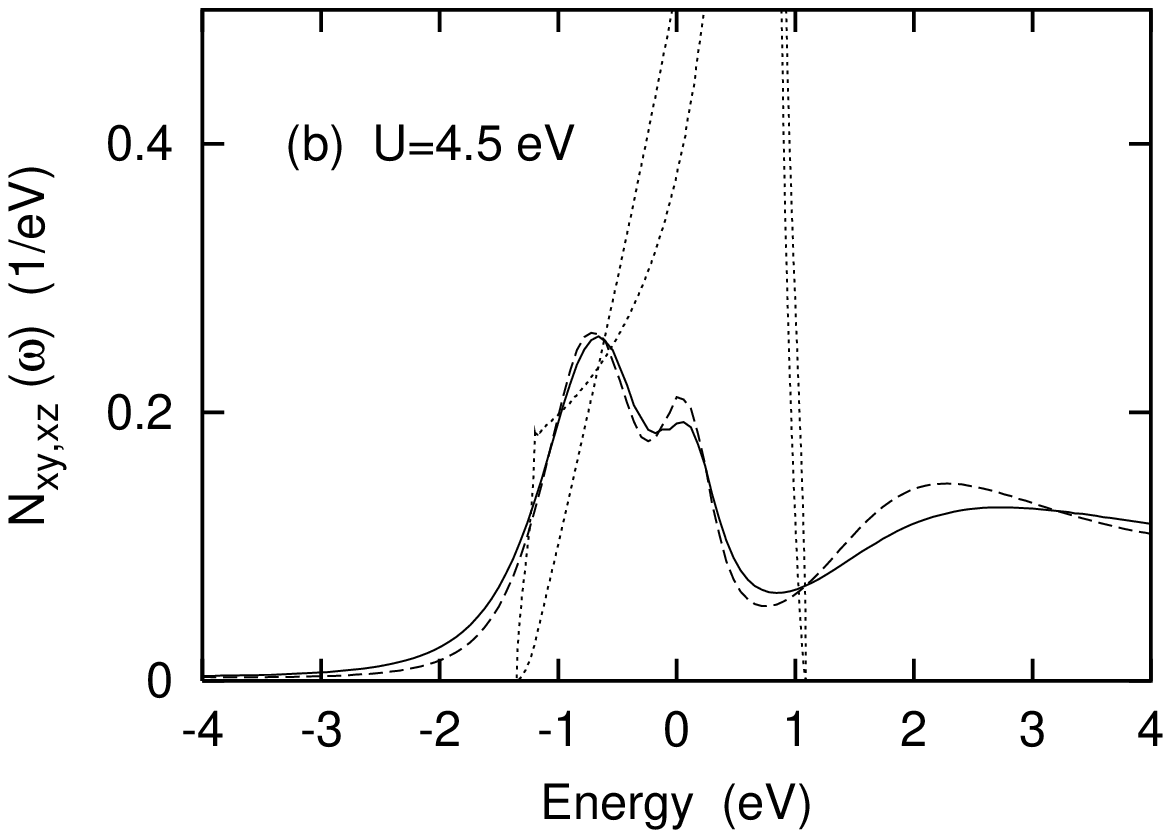}
    \includegraphics[height=7cm,width=4.0cm,angle=-90]{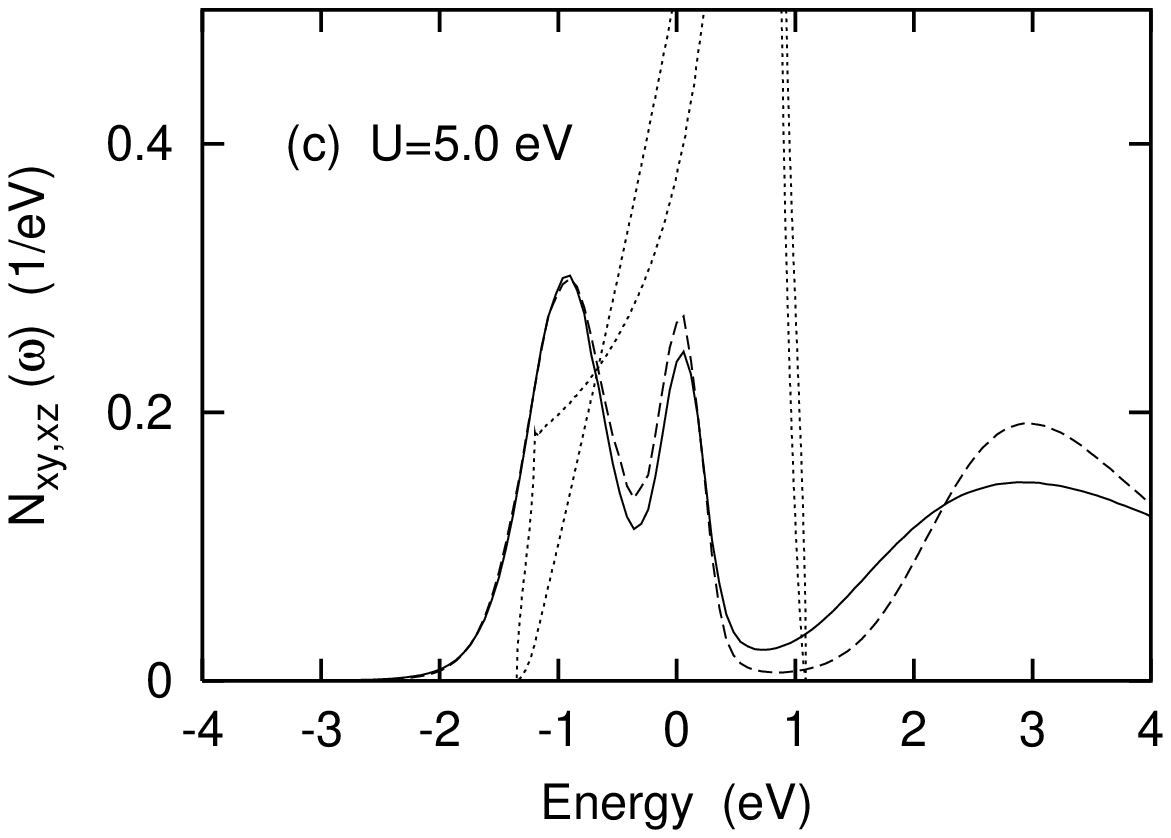}
\end{center}%vskip16cm
\caption{
Quasi-particle density of surface $d_{xy}$ states (solid curves) and 
$d_{xz,yz}$ states (dashed curves) of Ca$_{0.1}$La$_{0.9}$VO$_3$ 
($d^{1.9}$) for $U=4.0$, 4.5 and 5.0~eV derived from DMFT; $J=0.7$~eV. 
Dotted curves: bare densities of states.
}\end{figure}

According to these results we consider $x=0.1$ as the most promising
Ca concentration at which the bulk is already close to metallicity
while the surface layer might still be insulating. Fig.~3 
shows bulk quasi-particle spectra for Ca$_{0.1}$La$_{0.9}$VO$_3$ 
for various Coulomb energies in the critical region of the 
metal-insulator transition. The analogous surface spectra are 
shown in Fig.~4. All spectra exhibit the three-peak structure
characteristic of strongly correlated systems: a coherent peak
at $E_F$ and upper and lower Hubbard bands which appear as satellites
in photoemission or inverse photoemission data. As in the systems
studied previously\ \cite{lieprb}, as a consequence of the effective
narrowing of the $d_{xz,yz}$ subbands, correlations are more 
pronounced at the surface than in the bulk: the Hubbard peaks are
more intense and the coherent peak is a narrower. Since the 
QMC-DMFT calculations are performed at about 1450~K ($\beta=8$) 
the coherent peak persists even in the insulating phase. 
Nevertheless, in view of the temperature dependence of this feature
known from one-band systems\ \cite{georges,bulla} the spectra in 
shown in Figs.~3 and 4 suggest that at the surface the likelihood for 
a gap to open at $E_F$ at low temperatures is slightly greater than 
in the bulk. In the bulk there is clearly more spectral weight 
near  $E_F$ compared to pure LaVO$_3$ ($x=0$)\ \cite{lieprb},
reflecting the beginning metallic bulk behavior at $x=0.1$ Ca doping 
concentration. The surface spectra shown in Fig.~4, on the other hand,
indicate stronger tendency towards gap formation, in particular for 
$U=5$~eV.
 
In spite of the more pronounced surface correlations discussed above,
it is not clear whether extrapolation to low temperatures would yield
different Mott transitions for the bulk and the surface layer. Even 
though the spectral weight at $E_F$ is consistently smaller at the
surface, both $N_s(E_F)$ and $N_b(E_F)$ could still vanish at the 
same critical $U$. This would be analogous to the behavior found
in semi-infinite Heisenberg models: the magnetization near the 
surface is smaller than in the bulk, but the critical temperature
for the transition towards paramagnetism is the same\ \cite{magnetism}.
Only if surface and bulk are strongly decoupled their critical behavior
might be different. A careful examination of this question would also 
require to account for the coupling between the layer dependent 
`impurity' baths which we have neglected for computational reasons. 
Presumably this coupling whould diminish the differences between bulk 
and surface spectra.  

So far we have only considered the effect of band narrowing on the 
$d_{xz,yz}$ states which hybridize preferentially in planes normal 
to the surface. Surface reconstruction or distortion of oxygen 
octahedra presumably would reduce hopping also between $d_{xy}$ 
states in the first layer, thereby leading to additional enhancement 
of surface correlations. Moreover, less efficient electronic screening 
processes near the surface could increase the local Coulomb energy. 
Both effects would make the surface electronically different from the 
bulk. It is doubtful, however, whether this decoupling would be strong
enough to give rise to separate bulk/surface Mott transitions. 

In summary, we have calculated bulk and surface quasi-particle 
distributions for Ca$_{x}$La$_{1-x}$VO$_3$ using the dynamical mean 
field theory. At integer occupancy ($d^2$ for $x=0$) this perovskite 
material is a Mott insulator. At small Ca concentrations, however, the 
bulk material becomes metallic. As a result of the reduced atomic 
coordination at the surface, two of the $t_{2g}$ subbands exhibit 
appreciable band narrowing, giving rise to stronger correlation features 
in the quasi-particle spectra. In the range of critical local Coulomb 
interactions for the Ca concentration $x=0.1$ we have found at the
surface a tendency for the formation of an excitation gap at a lower 
value of $U$ than in the bulk. Whether this implies a coexistence of 
metallic bulk and insulating surface behavior is, however, difficult 
to estimate at present because of the finite temperature used in the 
QMC-DMFT calculations. Also, additional surface effects on the 
electronic structure such as reconstruction, the charge state of the 
surface layer, and weaker screening of the local Coulomb energy
should be considered.     

Acknowledgement: I like to thank A. Bringer, E. Eisenriegler, and 
D.D. Sarma for discussions. I also thank A.I. Lichtenstein for the 
QMC-DMFT code.

\end{document}